\documentclass[letterpaper,twocolumn,10pt]{article}
\usepackage{usenix2019_v3}

\usepackage{tikz}
\usepackage{amsmath}

\usepackage{filecontents}

\usepackage{color, colortbl}
\definecolor{Gray}{gray}{0.9}
\definecolor{LightCyan}{rgb}{0.88,1,1}
\definecolor{LightRed}{rgb}{1,0.88,0.88}
\definecolor{LightBlue}{rgb}{0.12,0.7,0.8}
\definecolor{LighterBlue}{rgb}{0.12,0.9,0.9}
\usepackage{pbox}
\usepackage{makecell}
\usepackage{amsmath}

\begin{document}

\date{}

\title{\Large \bf Toward Optimal Performance with Network Assisted TCP at Mobile Edge}

\author{
{\rm Soheil Abbasloo$^1$, Yang Xu$^2$, H. Jonathon Chao$^1$, Hang Shi$^3$, Ulas C. Kozat$^3$, Yinghua Ye$^3$}\\
New York University$^1$,  Fudan University$^2$, Futurewei Technologies$^3$
} 

\maketitle

\begin{abstract}
In contrast to the classic fashion for designing distributed end-to-end (e2e) TCP schemes for cellular networks (CN), we explore another design space by having the CN assist the task of the transport control. We show that in the emerging cellular architectures such as mobile/multi-access edge computing (MEC), where the servers are located close to the radio access network (RAN), significant improvements can be achieved by leveraging the nature of the logically centralized network measurements at the RAN and passing information such as its minimum e2e delay and access link capacity to each server. Particularly, a Network Assistance module (located at the mobile edge) will pair up with wireless scheduler to provide feedback information to each server and facilitate the task of congestion control. To that end, we present two \textit{\textbf{N}etwork \textbf{A}ssisted} schemes called NATCP (a clean-slate design replacing TCP at end-hosts) and NACubic (a backward compatible design requiring no change for TCP at end-hosts). Our preliminary evaluations using real cellular traces show that both schemes dramatically outperform existing schemes both in single-flow and multi-flow scenarios.
\end{abstract}

\section{Introduction}
\label{intro}
Various studies show that the general purpose TCP performs poorly in CNs~\cite{sprout, ex-tcp, verus, lte_depth, bufferbloat2, c2tcp} because of the key fundamental issue of the \textit{generalization}. Although the goal of having one general purpose TCP that can work in various networks is attractive, this goal made existing TCP a ``Jack of all trades, master of none''. A general purpose TCP requires to generalize the network and the reasons for congestion in the network. However, the generalization comes with the cost in performance. For instance, TCP Cubic (today's default TCP in most of the platforms) considers the loss of packets as a general indication of congestion. So, in the CN where stochastic packet losses exist, TCP Cubic misinterprets the stochastic packet losses as congestion which leads to its performance degradation. The generalization issue motivates a lot of recent congestion/flow control proposals to follow the \textit{domain-specific} design philosophy in which the design is limited to a specific network and leverages the characteristics of that network to boost the performance (e.g.~\cite{c2tcp, c2tcp2, sprout, verus} in CNs and ~\cite{dctcp, d3, hyline} in data center networks). 

On the other hand, although current distributed domain-specific TCPs can perform better than general TCPs in CNs, they are still suboptimal solutions. The reason is that, as proved by J. Jaffe~\cite{jaf}, no \textit{distributed} congestion control can converge to the operation point in which both the minimum delay and maximum throughput are achieved (Kleinrock's optimal point~\cite{optimal}). This result comes from the fact that by only using distributed e2e measurements (as in TCP), it will be unclear whether an increase in e2e delay measurements is due to another competing flow, a route change, or the client delaying the acknowledgment. The problem of distributed e2e measurements of TCP becomes even worse in CNs when downlink/uplink wireless scheduling delay, self-inflicted queuing delay, and channel capacity variations are considered.

Considering these shortcomings and the fact that new network latency and throughput requirements of the emerging applications such as real-time online gaming, automated vehicles, and virtual reality push the performance bar of TCP higher than the currently acceptable ones, we seek to boost the performance of TCP toward the optimal point in Mobile-Edge environment.

In CNs, there is usually rich information about each user equipment (UE) in a fine-grain timescale. For instance, base stations are aware of the history of throughput, per-UE queue occupancy, and the quality of channel of the UEs connected to them ~\cite{lte-book}. That motivated us to address the problem of distributed e2e measurements of TCP by investigating the benefits of leveraging the logically centralized network measurements and information gathering in CNs to assist the transport control at servers, especially when they are located at the edge of the network.

To that end, we introduce a Network Assistance (NetAssist) entity at the mobile edge which periodically sends the bottleneck link's (BL) bandwidth (Bw) and delay information to end-hosts in an out-of-band (OoB) manner. As a proof of concept, we present two schemes called NATCP and NACubic. NATCP is a clean-slate design replacing traditional TCP at end-hosts using feedback information from NetAssist and calculate congestion window (Cwnd) and pacing rate. However, NACubic is a backward compatible design requiring no change in traditional TCP at end-hosts. It simply works on top of the current TCP and only caps Cwnd and pacing rate to the values calculated using feedback information. Using real cellular traces, we evaluated NATCP and NACubic and compared them with various state-of-the-art TCP schemes (detailed in section~\ref{eval}). We showed that both NATCP and NACubic significantly outperform other schemes and achieve higher power (defined as $\frac{throughput}{delay}$) and lower 95th percentile delays. For instance, NATCP and NACubic achieve more than 2.7$\times$ higher power compared to ABC~\cite{abc} (the most recent ECN-based approach in CNs). 
\section{Design}
The golden three goals of any congestion control are to achieve: 1) maximum throughput, 2) minimum delay, and 3) fairness among flows. However, none of the existing TCP proposals in the literature achieves all these 3 goals simultaneously. For instance, most of the TCP proposals use the AIMD (additive increase multiplicative decrease)~\cite{aimd} algorithm to increase their fairness performance among multiple flows. However, this affects the two other performance metrics by slowing down the Cwnd growth in the case of a sudden increase in available bandwidth and harsh Cwnd reduction in the case of a sudden decrease in available bandwidth. Therefore, to achieve all 3 goals, we decouple the problem of achieving fairness from the problem of achieving maximum bandwidth and minimum delay. 
\subsection{Fairness and the Role of UE}
We divide the fairness problem into two sub-problems: 1) fairness among different UEs (Inter-Fairness) and 2) fairness among the flows of one UE (Intra-Fairness).

\begin{figure}[!t]
\centering
\includegraphics[width=0.8\linewidth,height=0.8in]{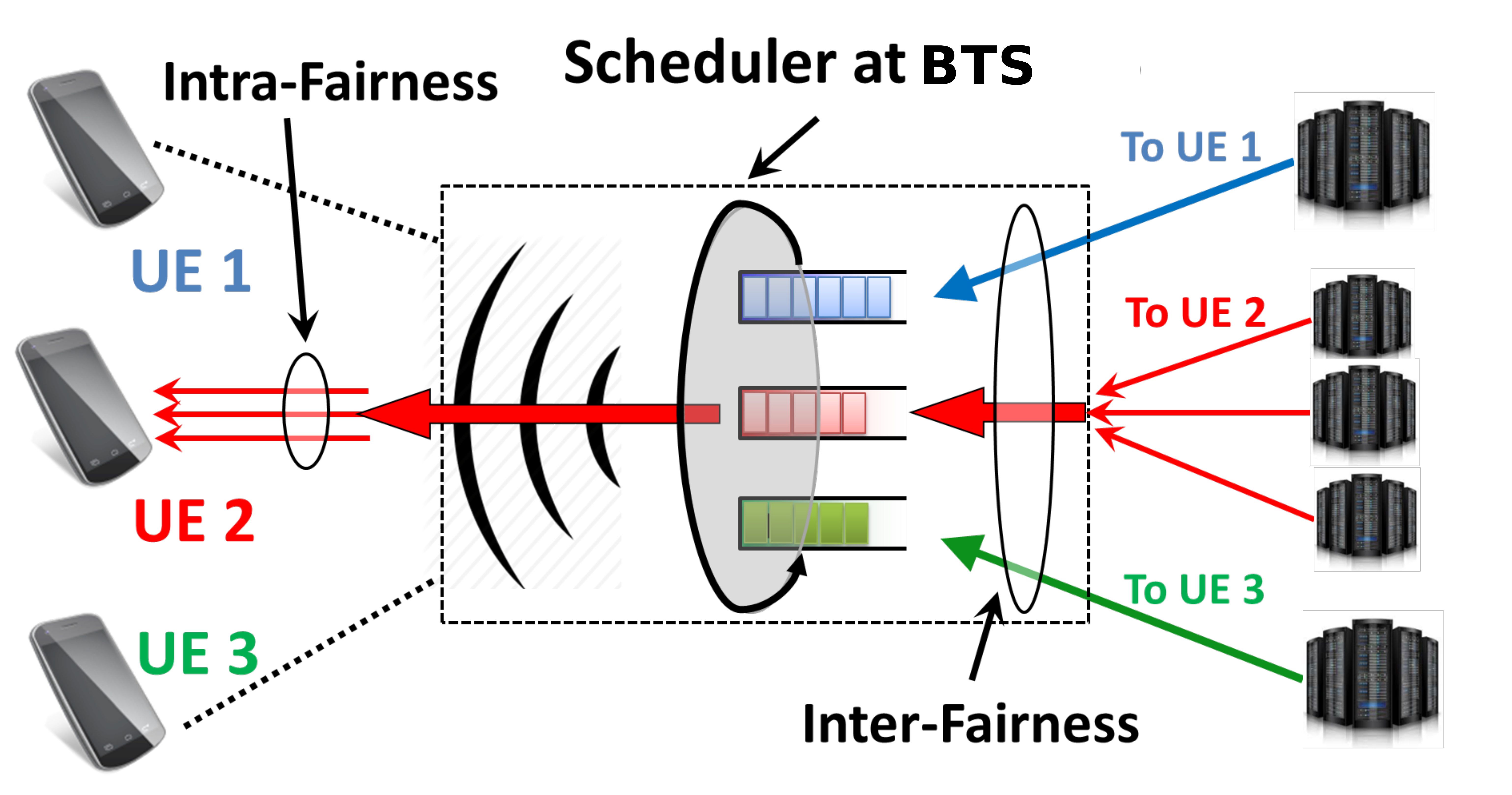}
\caption{Scheduler and per-UE queues at BTS}\label{fig_sch}
\end{figure}

In contrast to wired networks, there are per-user queues and there is a scheduler in CNs (Figure~\ref{fig_sch}). The scheduler is responsible for scheduling downlink/uplink per-user queues and bringing the fairness among different UEs at base transceiver station (BTS)~\cite{lte-book}. In other words, at the bottleneck switch in a wired scenario, an aggressive TCP can fill up the buffer and take the bandwidth from other TCPs; however, in cellular scenario, scheduler at BTS schedules the access of the aggressive TCP to the wireless channel so that all flows can have a fair amount of wireless link's bandwidth regardless of the TCP protocol that they use. Therefore, the inter-fairness problem has already been resolved by the BTS' scheduler. However, there is still the problem of intra-fairness. 

In a general network, the number of competing flows at the bottleneck queue is not known. Therefore, in a classical distributed congestion control approach, each end-host tries to independently adjust its sending rate using techniques such as AIMD to give room to other likely competitors in the network. However, in CNs, due to the existence of per-UE queues, the number of competing flows at each queue at BTS is known by each UE (it is equal to the flows destined to the UE). Considering that, there are two approaches to resolve the Intra-fairness in CNs: 1) Since TCP is a closed-loop algorithm and a sender always requires to receive acknowledgments from the receiver to send more packets, the receiver (UE) who knows the competing flows can simply regulate the sending rates of the senders by regulating the sending rates of their corresponding Ack packets. 2) UE can directly inform servers about the competing flows (destined to that UE) and help them to adjust their sending rates accordingly by sending fairness information feedback to them\footnote{In the simplest case, fairness information is the number of competing flows at UE, while a much richer feedback can be constructed considering a more sophisticated definition of fairness.}.

In our design, each UE reports the fairness information to the servers using the Ack packets (Step \# 1 in Figure~\ref{fig_big}) either indirectly (first mentioned method) or directly (second described method).
\subsection{Towards the Optimal Point}
The notion of having a logically centralized entity providing guidance/control to/over UEs is already part of the CN. For instance, BTS dynamically schedules the transmission/reception of packets to/from UEs and periodically sends control messages to UEs indicating their scheduled time slots. Intuitively, NATCP exploits the same idea but this time for providing assistance to the servers in CN aiming to operate close to Kleinrock's optimal point~\cite{optimal}.
\subsubsection{NetAssist}
\begin{figure}[!t]
\centering
\includegraphics[width=0.85\linewidth,height=1in]{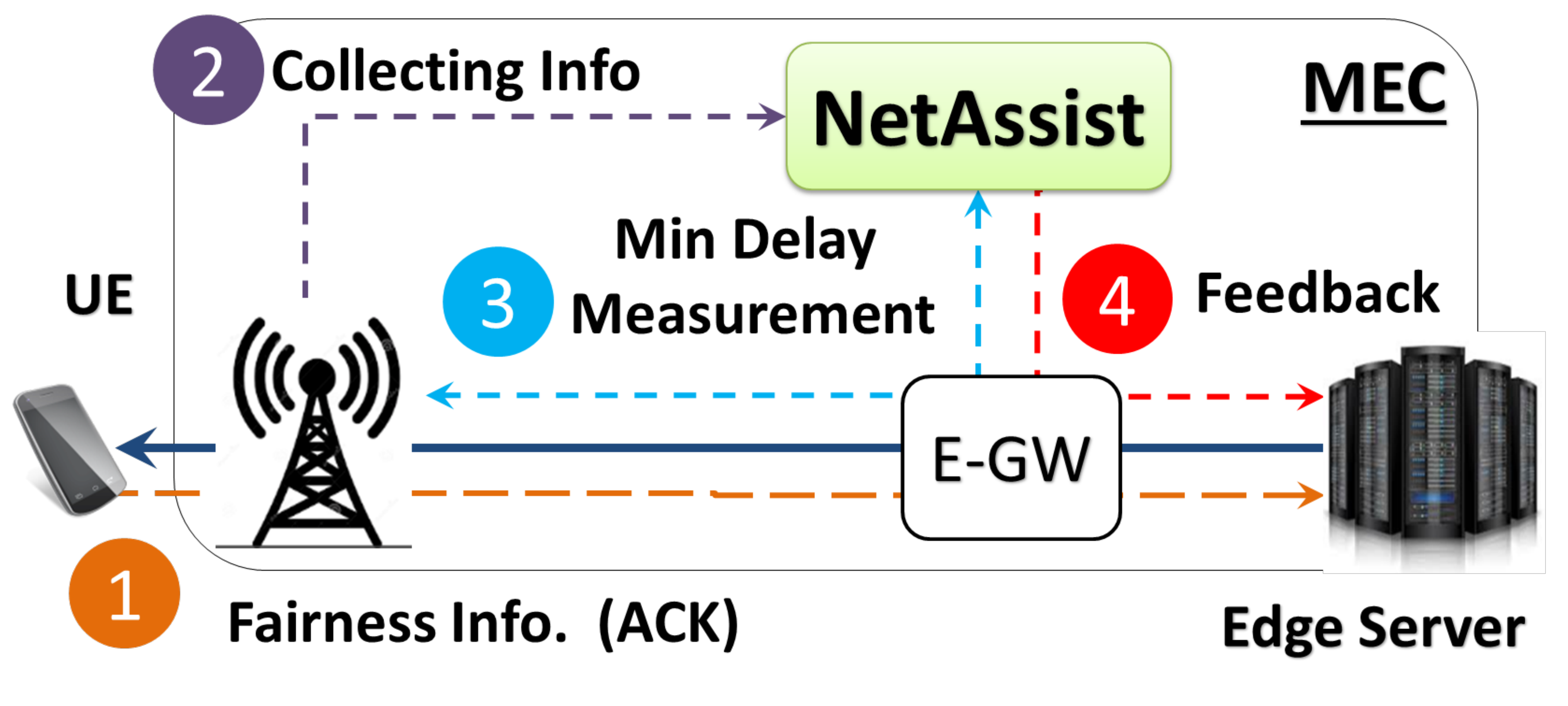}
\caption{The big picture of NATCP/NACubic structure}\label{fig_big}
\end{figure}

Placing customary servers inside the mobile operator's networks reduces the intrinsic e2e delay. However, due to the wireless nature of cellular access links which causes multiple order of magnitude fast capacity fluctuations, cellular access links (known as last-mile) remain the main BL in CNs. To estimate the BL's bandwidth and the minimum delay, we propose that the end-hosts get help from the network itself. In particular, we introduce a new entity called Network Assistance (NetAssist) in the network which periodically collects information regarding the BL's bandwidth and the minimum delay for each UE (Steps \#2 and \#3 in Figure~\ref{fig_big}), and sends digested feedback (Step \#4 in Figure~\ref{fig_big}) to the servers, that have registered to get congestion assistance service from the CN, in OoB manner. This feedback will be used by servers to set Cwnd and pacing rate and send data accordingly.

\textbf{BL's bandwidth ($BL_{Bw}$):} The $BL_{Bw}$ (which is already available at BTS in a fine-grained timescale) will be averaged at each feedback period and sent by NetAssist. Our sensitivity analysis indicates that a period of $50ms$ for sending feedback from NetAssist to the server gives reasonable results. With feedback periods much larger than 50ms, the server misses channel fluctuations and cause NATCP/NACubic to react too slowly to them. On the other hand, feedback periods much smaller than 50ms may not give enough time for the server to react against the highly dynamic variation of cellular access link capacity.

\textbf{minRTT:} minRTT is the minimum RTT that a packet in the network can experience excluding any waiting time in the queues and sender/receiver response time delays. To measure minRTT, we segment it into 3 parts: 1) The minimum delay of the network (from edge gateway (E-GW) to BTS and vice versa), 2) The transmission delay and scheduling delay of packets at BTS in the downlink direction, and 3) The transmission delay and scheduling delay of packets at BTS in upstream direction\footnote{due to the server proximity in MEC architecture, the latency from the server to E-GW accounts for a small portion of the overall latency, so it can be ignored. Also, we ignore the intrinsic wireless signal propagation delay between BTS and UE which is usually very small}. 
To measure the first part, a simple way is to use higher priority probing packets that are periodically sent to the BTS from NetAssist (or a dedicated monitoring module). Due to their higher priorities, probing packets will not experience any queuing delay and will show the delay excluding queuing delays at the network. Also, delay measurements can be done by using more advanced approaches provided by programmable devices such as attaching and stripping meta-data information to data packets (e.g., ~\cite{inband}) (This part does not need to be measured per server, because the network's minimum delay is roughly the same for all of them). 

Since the sending/receiving rate of packets to/from the UE is controlled by the wireless scheduler, the depletion/reception rate of the UE at BTS in a time interval declares the sum of averaged scheduling delay and averaged transmission delay. Therefore, to calculate parts 2 and 3, NetAssits uses averaged sending/receiving rates to/from UEs at BTS over the feedback time period. NetAssist calculates minRTT by adding these 3 parts and sends it as part of feedback information to the servers.

\subsubsection{Logic at the Edge Server}
Getting feedback information from the NetAssist and fairness information from the UE, the server will calculate Cwnd and pacing rate for flows. Here, we propose two solutions. One is a clean-slate design which completely replaces TCP at servers (called NATCP) and the other one is a backward compatible design which without modifying the existing TCP at server (e.g., Cubic~\cite{cubic}), applies received feedback from NetAssist (So, without loss of generality, we assume the default TCP is Cubic, and we call the second scheme NACubic). 
\\\textbf{NATCP}: 
In NATCP, servers use $BL_{Bw}$ to set the pacing rate of the flow's outgoing packets (conceptually pushing the BL to the server). In addition, servers calculate Cwnd using Equation~\ref{eq_cwnd} where minRTT and $BL_{Bw}$ are the minimum delay and BL bandwidth feedback received from NetAssist, respectively and $\beta$ is the fairness coefficient (received from the UE). At source/destination packets will usually be queued between different network layers at OS. Therefore, Cwnd value in the TCP layer does not necessarily equal to the number of in-flight packets. In Equation~\ref{eq_cwnd}, $\alpha$ represents a scaling factor to compensate for the difference of Cwnd with the real in-flight packets. We set $\alpha$ to 2 throughout this paper.
\begin{equation} 
\label{eq_cwnd} 
Cwnd=\alpha\times \frac{1}{\beta} \times minRTT \times BL_{Bw}
\end{equation} 
\textbf{NACubic}: In NACubic, the server does not directly use feedback information to calculate the Cwnd. Instead, it simply caps the Cwnd calculated by Cubic (or other TCP variants) to the MaxCwnd calculated by Equation~\ref{eq_cwnd}. In this way, there is no need to modify the TCP at the server. Similar to NATCP, NACubic sets pacing rate to $BL_{Bw}$\footnote{Setting per-flow pacing rate functionality is currently part of the TCP stack in Linux kernel 4.13.}.

\subsection{Why it works}
\textbf{Is Centralized Delay Feedback Really Necessary?} To shed light on why we need both centralized BW and delay measurements, we use one trace from prior work\cite{sprout} (namely Verizon-EVDO-downlink) and send traffic from the server to UE following instructions detailed in section~\ref{eval}. 
\begin{figure}[!t]
    \centering
    \includegraphics[width=.85\linewidth,height=0.9in]{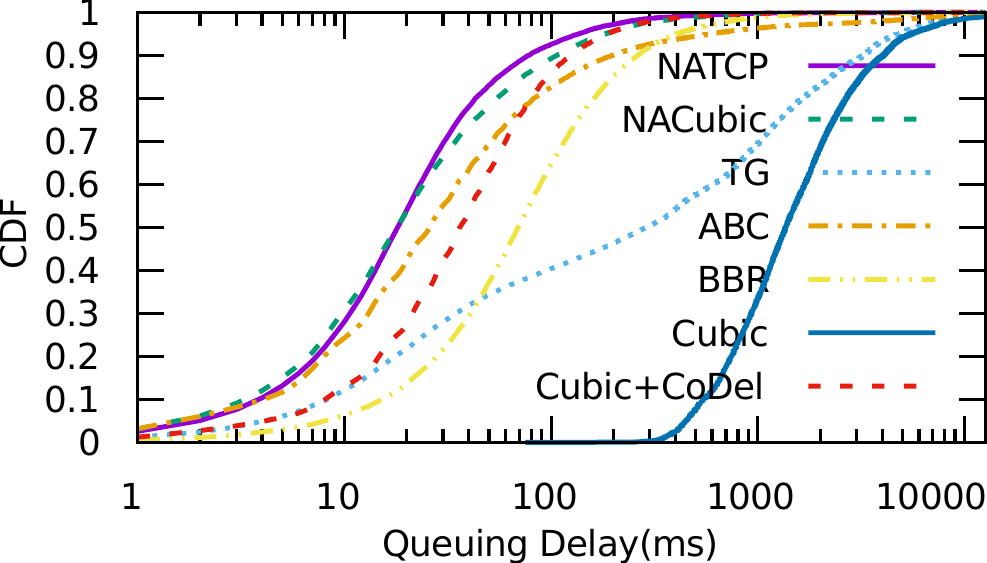}
    \caption{CDF of queuing delay over Verizon-EVDO.downlink trace}
    \label{fig_cdf-evdo}
\end{figure}
In the CN where BL BW is highly dynamic, current distributed e2e estimations of minRTT at end-hosts might not be likely valid for the next following RTTs. This issue escalates when the intrinsic e2e delay (excluding queuing delay) of the CN is very low. For instance, consider a network with the intrinsic low e2e delay of 10ms (as for the network of this section's experiment). When link capacity is $1.2 Mbps$ at arbitrary time $t$, the transmission delay for HOL (head-of-line) packet $p$ of size $1500B$ becomes $10ms$ alone. This also makes $10ms$ increase in the sojourn time (waiting time) of all packets queued behind packet $p$. So, for networks similar to MEC-based CNs (with intrinsic low e2e delay and highly variable BL (last-mile)), the minimum RTT will be significantly impacted by the queuing delay (including transmission time and sojourn time, either because of wireless scheduling delay or variation of the transmission times) in small time scales. Therefore, the distributed e2e delay estimations that require at least a few RTTs will not provide a good estimation of the real minimum RTT of the current time. 

Fig.\ref{fig_cdf-evdo} shows the impact of this phenomena. When link capacity is highly dynamic (as in Verizon-EVDO trace), TG scheme, which uses only BW feedback and counts on RTT measurements at the server, performs very poorly. Also, BBR which estimates minRTT in a distributed manner performs poorly, especially in high percentile regions. ABR uses ECN bits set by BTS in the downlink direction and piggybacked by Acks to the server to adjust the rate quickly. However, due to the high variations of the queuing delay, even this ECN-based scheme, as Fig.\ref{fig_cdf-evdo} illustrates, cannot perform very well. This clearly shows the importance and advantage of using centralized delay and centralized BW measurements/feedback in our design. 

\textbf{OoB Signaling:}
Table~\ref{table_inband} compares the improvements of Power$_{95}$ (defined as throughput /95th\ \%tile\ queuing\ delay) gained by using different feedback info (only $BL_{Bw}$ vs. $BL_{Bw}$ and minRTT(NATCP)) and different way of sending that (OoB vs. IB (In-Band)) for the setup detailed in section~\ref{eval} (using ATT-LTE-2016 traces~\cite{mahi})\footnote{Here, feedback was generated every 10ms.}.
\begin{table}[!h] \renewcommand{\arraystretch}{1} \caption{Normalized $P_{95}$ for various schemes} 
\label{table_inband}
\scriptsize
\centering
    \begin{tabular}{c|c|c|c|c}
    Scheme& \cellcolor{LightRed} Bw (IB) & Bw (OoB)& NATCP (IB)& \cellcolor{LighterBlue}NATCP (OoB)\\
    \hline
    $P_{95}$& \cellcolor{LightRed}1$\times$ &3.47$\times$ & 7.32$\times$ & \cellcolor{LighterBlue}9.31$\times$ \\
    \end{tabular}
\end{table}
Generally, OoB signaling performs better than IB signaling, because feedback will not be delayed due to downlink/uplink scheduling and queuing delays. Hence, NetAssist periodically sends feedback to servers in an OoB manner as opposed to IB signaling (e.g., as in~\cite{tg}), though it might add small signaling overhead (check section~\ref{sec_discussion}).
\subsection{Discussion}
\textbf{Signaling Overhead:}
\label{sec_discussion}
NetAssist only sends two values as feedback, so let us assume the feedback packet size to be 64Bytes. Hence, the data rate for the OoB feedback per UE will be $\frac{64B}{50ms}=10kbps$. Even, when the number of users is 1,000 (per NetAssist), there will only be a total of 10Mbps signaling overhead. This is quite acceptable considering current LTE networks where each UE can reach the rates of tens of Mbps and the fact that next-generation 5G CNs aim to provide hundreds of Mbps per UE~\cite{5g_1}~\cite{5g_2}.

\textbf{Is NATCP/NACubic scalable?} 
NetAssist module pairs up with BTS to gather information without altering or interfering with the scheduler's operations. The NetAssist pairing up with a certain BTS works independent of the NetAssist module paring up with another BTS, because there is no need to pass information from one NetAssist to another one. For instance, in the case of handover, when a user disconnects from one BTS and connects to a new BTS, no context-switching happens. The scheduler in the new BTS does not get throughput history of the new user from previous BTS (to schedule its transmission).\footnote{This behavior has been confirmed through our private discussions with CN provider engineers} That is the same for NetAssist. In other words, there is no need for having either a global view or previous states of the whole CN in NetAssist. These properties render NATCP/NACubic a highly scalable design.

\textbf{Deployment of NetAssist:} The traditional issue with the approaches requiring changes in the network is the feasibility and the high costs of those changes for the network operators. However, the customizable nature of recent software-based trends such as function virtualization~\cite{nfv} makes the deployment of new services in the network feasible while having low costs. For instance, NetAssist module can be realized as a virtual function in C-RAN (cloud/centralized RAN)\cite{cran} and network operators can use the available pool of servers at MEC to make the virtual NetAssist modules. 

\textbf{What if network feedback does not reach the server?} We can use a simple timer concept to monitor loss of feedback information. In other words, if feedback information is not received within a certain time, the timer will timeout and NATCP/NACubic will revert to normal TCP at the sender. 


\section{Preliminary Evaluations}
\label{eval}
We have implemented NATCP/NACubic in Linux Kernel 4.13, written proof of concept NetAssist modules in user-space, and compared performances of NATCP and NACubic with existing schemes using Mahimahi~\cite{mahi} (Available at: \url{https://github.com/Soheil-ab/natcp}).

\textbf{Cellular Traces}: 
We use 16 different cellular traces collected in prior work (\cite{mahi} and \cite{sprout}) as the cellular link traces.

\textbf{Topology and Metrics}:\label{sec_topology} We use a server which is connected to a client through Mahimahi emulator. Mahimahi emulates the cellular access link using the given cellular traces.  To have an MEC like network, we set the minimum RTT to 10ms. The per-user buffer size at BTS is set to 150KB\footnote{Although the specific buffer size at BTS for each UE is not in public domain, we have tried to select the buffer size by comparing results from emulations with results from the real-world experiments for Cubic.}. The minimum RTT from NetAssist to the server is set to 2ms\footnote{Due to the server proximity, this delay in practice is a few hundreds~of~us.}. We start a flow between a server and a client that lasts for the entire duration of each trace and measure the average and 95th percentile per packet queuing delay, average throughput, power (defined as $\frac{Throughput}{Avg.\ queuing\ delay}$), and power$_{95}$ (defined as $\frac{Throughput}{95th\ \%tile\ queuing\ delay}$).

\textbf{Schemes Compared}: We compare NATCP and NACubic with three e2e TCP designs: Cubic~\cite{cubic}, BBR \cite{bbr}, PCC-Vivace~\cite{vivace}; four e2e cellular schemes: C2TCP~\cite{c2tcp}, Verus~\cite{verus}, Sprout~\cite{sprout}, and Westwood~\cite{west}; one ECN-based scheme: ABC~\cite{abc}, one feedback-based scheme: TG~\cite{tg}; and one AQM design Cubic+Codel~\cite{codel}. We tried to tune different schemes to have their (on average) best performance, though our tuning might not be perfect.
For the TG scheme, we use the same BW feedback used by NATCP which represents averaged available bandwidth on 50ms time windows.

\textbf{Overall Results}:
\begin{figure}[!t]
\centering
        \includegraphics[width=\linewidth,height=1.3in]{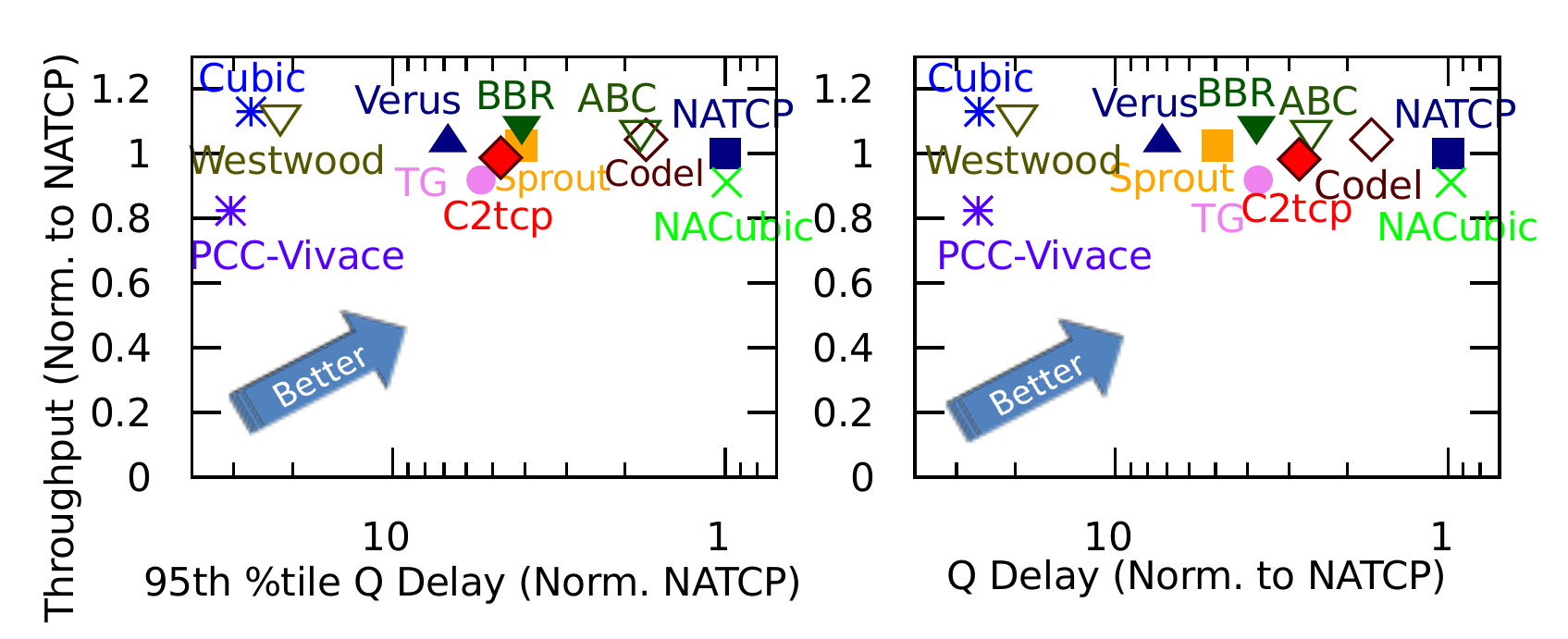}
        \caption{Normalized results to NATCP averaged over all traces}\label{fig_overall}
\end{figure}
For each trace, we normalize the performance results of different schemes to the NATCP's performance for that trace. Later, we average all normalized results over all 16 cellular traces from \cite{sprout} and \cite{mahi}. Figure~\ref{fig_overall} shows these overall results. As these graphs illustrate, the distinguishing performance difference among schemes is the delay responses of them. NATCP and NACubic outperform all other schemes in terms of delay performance, while compared to Cubic (which gets the highest throughput among all schemes), they only compromise a bit of throughput (7\% and 20\% throughput reduction respectively). Also, Table~\ref{table_power} summarizes the average relative Power and Power$_{95}$ improvement for NATCP compared with other schemes, averaged over all CNs traces. All e2e schemes suffer from the distributed measurements issue leading to their poor performance. On the other hand, NATCP beats TG (which uses throughput guidance from the network but still leaves the delay estimations to the end-host) and achieves 3.86$\times$ better power. This clearly shows the advantages of having centralized delay measurements over the distributed measurements. 
\begin{table}[!t] \renewcommand{\arraystretch}{1} \caption{Power reduction averaged across all traces} 
\label{table_power}
    \scriptsize
    \begin{minipage}[pt]{0.39\linewidth}
     \centering
            \begin{tabular}{|c|c|c|}
            \hline
            Scheme& P & P$_{95}$\\
            \hline
            \cellcolor{LightBlue}NATCP& \cellcolor{LightBlue}1$\times$ & \cellcolor{LightBlue}1$\times$ \\
            \cellcolor{LighterBlue}NACubic& \cellcolor{LighterBlue}1.09$\times$ & \cellcolor{LighterBlue}1.1$\times$ \\
            ABC& 2.71$\times$ & 1.83$\times$ \\
            C2TCP&2.79$\times$ & 4.83$\times$ \\
            BBR& 3.46$\times$ & 3.76$\times$ \\
            TG& 3.86$\times$ & 5.65$\times$ \\
            \hline
            \end{tabular}
            \end{minipage}
            \hfill
            \begin{minipage}[pt]{0.58\linewidth}
     \centering
            \begin{tabular}{|c|c|c|}
            \hline
            Scheme& P & P$_{95}$\\
            \hline
            Cubic+Codel& 1.63$\times$ & 1.66$\times$\\
            Sprout& 4.8$\times$ & 3.98$\times$\\            
            Verus& 7.01$\times$ & 6.67$\times$ \\
            Cubic& 22.85$\times$ & 23.85$\times$\\
            Westwood& 17.68$\times$ & 19.60$\times$\\
            PCC-Vivace& 28.61$\times$ & 35.54$\times$\\            
            \hline
            \end{tabular}
            \end{minipage}
\end{table}

\begin{figure}[!t]
\centering
    \begin{minipage}[pt]{0.39\linewidth}
            \scriptsize
            \begin{tabular}[width=\textwidth,height=0.9in]{|c|c|} \hline
            Scheme& \pbox{20cm}{Normalized \\ Number of \\ReTrans.}\\ \hline
            \cellcolor{LightBlue}NACubic& \cellcolor{LightBlue}1$\times$ \\
            \cellcolor{LighterBlue}NATCP& \cellcolor{LighterBlue}4$\times$ \\
            Cubic& 20$\times$\\
            BBR& 25$\times$ \\
            Cubic+Codel& 54$\times$\\
            TG& 1470$\times$\\ \hline
            \end{tabular}
            \hfill
    \end{minipage}
    \hfill
    \begin{minipage}[c]{0.55\linewidth}
        \includegraphics[width=0.9\textwidth,height=1in]{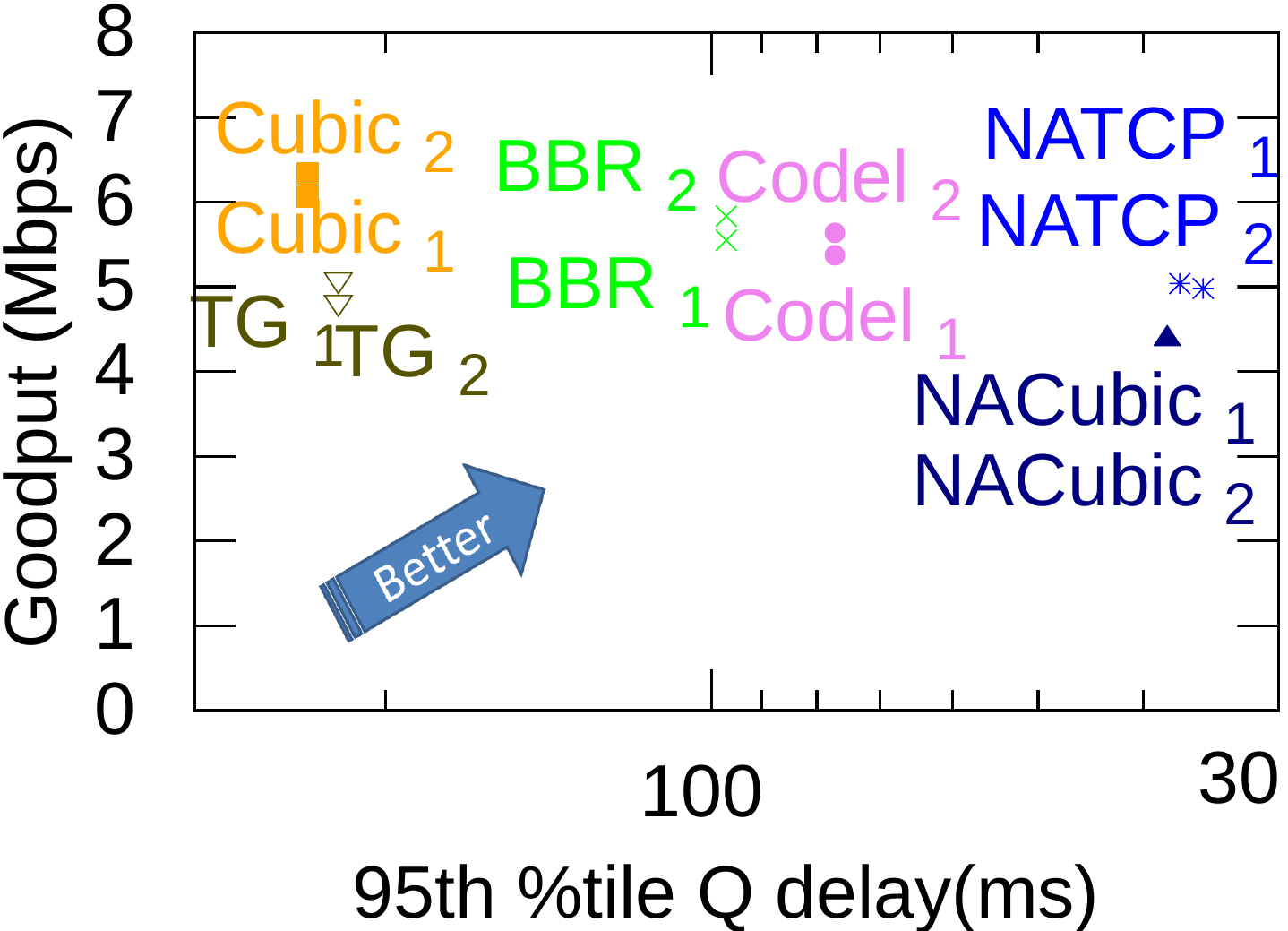}
    \end{minipage}
    \caption{The 95th \%tile Q delay, average Goodput, and total number of retransmissions for 2 flows competing at UE}
    \label{fig_fairness}
\end{figure}
\textbf{Intra-Fairness}:
To evaluate intra-fairness of NATCP/NACubic, we send traffic from two servers to one UE and report the goodput and number of packet retransmissions for each flow (we use TMobile-LTE-driving~\cite{mahi} as our access link). 
Figure~\ref{fig_fairness} shows the results. NATCP and NACubic significantly outperform other schemes in terms of achieving fair throughput without compromising the delay. For instance, compared to Codel, the state-of-the-art delay-based AQM design, both NATCP and NACubic flows achieve 2$\times$ lower 95th percentile delay. Also, NACubic and NATCP achieve the first and second minimum number of packet retransmissions among all schemes. For instance, compared to Codel, NACubic reduces packet retransmissions more than 50$\times$. 
The poor performance of TG comes from the fact that it has no mechanism to resolve fairness and it leaves delay measurements to the end-host. 

\section{Conclusion}
We have shown that significant improvements can be achieved by moving from a classical distributed e2e TCP to a Network Assisted TCP in MEC-based CNs. Network Assisted TCP exploits the logically centralized network measurements in CNs where the notion of centralized control and information gathering is a common practice. Based on that, we have presented two schemes called NATCP (a clean-slate design) and NACubic (a backward compatible design). Our preliminary results show that both schemes dramatically outperform prior work in terms of power in both single-flow and multi-flow scenarios. 
\section{Discussion Topics}

\textbf{ The controversial points of the paper}: The key controversial point of NATCP is its novel design philosophy which contradicts the design of fully distributed e2e TCPs. We show that the key architectural advances such as MEC can be leveraged to greatly improve the performance of TCP by going from a fully distributed e2e design to a network-assisted design at Mobile Edge. 

\textbf{The type of discussion this paper is likely to generate in a workshop format}: On the one hand, we hope that this work will generate discussions about reconsidering TCP design approaches among protocol designers and motivates them to exploit network-assisted framework for TCP in CNs. On the other hand, we hope that the encouraging preliminary results presented here show an opportunity for providing such novel and unconventional services to end-hosts and convince people that it is a viable solution to the CN. We believe that more work in this direction will be generated in the future. 

\textbf{What kind of feedback we are looking to receive}: Any feedback on the implementation of the proposed network-assisted framework is highly appreciated.

\textbf{Under what circumstances the whole idea might fall apart}: In scenarios where there is no feedback because the NetAssist and end-hosts are disconnected, the NATCP will roll back to the default TCP (without using the feedback). Therefore, circumstances representing scenarios without any feedback will not cause issues. However, if components dealing with the measurements (such as BTS) are providing wrong feedback, that might cause a problem. Those circumstances should be handled by extra safety mechanisms to always be sure that the measurement information provided to the NetAssist is correct.

\textbf{The open issues the paper does not address}: Although the congestion information exposed to servers will not cause security breaches, the communication among various entities need to be secured. So, one open issue that is not addressed in our paper is how to provide security for communications among different entities. In addition, the detailed protocols for communication among various components including NetAssist and end-hosts need to be investigated and standardized. 

\bibliographystyle{plain}

\end{document}